\documentclass[prd,showpacs,amsmath,amssymb,superscriptaddress,preprintnumbers,nofootinbib,showkeys]{revtex4}

\begin{document}

\newcommand{\nablab}{{\mathop {\rule{0pt}{0pt}{\nabla}}\limits^{\bot}}\rule{0pt}{0pt}}

\newcommand{\Jtr}{{\mathop {\rule{0pt}{0pt}{J}}\limits^{\bot}}\rule{0pt}{0pt}}

\title{Relativistic nonlinear axion magnetohydrodynamics}

\author{Timur Yu. Alpin}
\email{Timur.Alpin@kpfu.ru} \affiliation{Department of
General Relativity and Gravitation, Institute of Physics, Kazan
Federal University, Kremlevskaya str. 16a, Kazan 420008, Russia}

\author{Alexander B. Balakin}
\email{Alexander.Balakin@kpfu.ru} \affiliation{Department of
General Relativity and Gravitation, Institute of Physics, Kazan
Federal University, Kremlevskaya str. 16a, Kazan 420008, Russia}

\author{Alexei V.  Vorohov}
\email{avvorohov@stud.kpfu.ru} \affiliation{Department of
General Relativity and Gravitation, Institute of Physics, Kazan
Federal University, Kremlevskaya str. 16a, Kazan 420008, Russia}

\date{\today}

\begin{abstract}
The new nonlinear axionically extended version of the general relativistic magnetohydrodynamics is formulated. The self-consistent formalism of this theory is based on the introduction into the Lagrangian of the new unified scalar invariant, which is quadratic in the Maxwell tensor, and contains two periodic functions of the pseudoscalar (axion) field. The constructed unified invariant and the elaborated nonlinear theory as a whole, are invariant with respect to two symmetries: first, the discrete symmetry associated with the properties of the axion field; second, the Jackson's SO(2) type symmetry intrinsic for the electromagnetism. The subsystem of the master equations, which describes the velocity four-vector of the hydrodynamic flow, is constructed in the framework of Eckart's theory of viscous heat-conducting fluid. The axionically extended nonlinear Faraday, Gauss and Ampere equations are supplemented by the ansatz about the large electric conductivity of the medium, which is usually associated with vanishing of the electric field. We have suggested two  essentially new nonlinear models, in the framework of which the anomalous electric conductivity is being compensated by the appropriate behavior of the finite pseudoscalar (axion) field, providing the electric field in the magnetohydrodynamic flow to be finite (either to be proportional to the magnetic field, or to the angular velocity of the medium rotation).
\end{abstract}
\maketitle

\section{Introduction}

Relativistic magnetohydrodynamics is the important part of the modern cosmic plasma physics and fluid dynamics in the context of applications to
the theory of matter accretion to the rotating neutron stars and black holes,  to the theory of structure of magnetospheres of pulsars and Sun (see, e.g., \cite{L1}-\cite{MHD8} for references and description of the main problems). The canonic formalism of relativistic magnetohydrodynamics, described in the famous book of Andr\'e Lichnerowicz \cite{L1}, is constantly being extended to solve new astrophysical and cosmological problems. For instance, in the excellent paper of Massimo Giovannini  \cite{MHD4} the reader can find the applications of anomalous magnetohydrodynamics to the relativistic domains with extreme characteristics. In that paper also one can find a new element of fluid dynamics, namely, the theory of scalar/pseudoscalar fields interacting with electromagnetic fields.  In the paper \cite{MHD2} the authors consider fluids with chiral properties. The interest to these problems is not accidental, in fact, we are on the threshold of formulation and active use of the relativistic axion magnetohydrodynamics, which deals with interaction of the cosmic axionic dark matter with magnetohydrodynamic flows.

We consider the relativistic axion magnetohydrodynamics as an essential part of the relativistic theory of the axionically active plasma (see, e.g., \cite{AAP1} - \cite{AAP5} for some specific results for such plasma). We have introduced two new elements to the theory of axionically active systems. The first element is connected with the nonlinear approach to the description of the axion - photon coupling. The standard idea is to introduce into the Lagrangian the term $\frac14 \phi F^{*}_{mn}F^{mn}$ as it was done by the pioneers of the axion physics \cite{ax1} - \cite{ax9}. The pseudoscalar (axion) field $\phi$ enters this term linearly in front of the pseudo-invariant of the electromagnetic field presented by the convolution of the Maxwell tensor $F_{mn}$ and its dual $F^{*}_{mn}$. This term is invariant with respect to the discrete symmetry transformation $\phi \to \phi + 2\pi n$ ($n$ is an integer), since the rest term $2\pi n F^{*}_{mn}F^{mn}$ is the perfect divergence and thus it can be avoided from the action functional. If one uses arbitrary nonlinear function $f(\phi)$ instead of linear function $\phi$, this symmetry happens to be lost. Clearly, this function has to be periodic $f(\phi{+}2\pi n)= f(\phi)$, odd, and has to tend to $\phi$, when $\phi$ is small. One can choose, for instance, $f(\phi) = \sin{\phi}$. But we went further and applied the Jackson's SO(2) symmetry intrinsic for the electrodynamics \cite{Jackson}, thus obtaining the new unified invariant term
\begin{equation}
{\cal I} = \frac14 \left(\cos{\phi} F_{mn}F^{mn} + \sin{\phi} F^{*}_{mn}F^{mn} \right) \,,
\label{000}
\end{equation}
with necessary periodicity \cite{B1,B2}. When $\phi=0$, we deal with the standard Lagrangian of the electromagnetic field $\frac14 F_{mn}F^{mn}$; when $\phi$ is small, the new linear term is
$\frac14 \phi F^{*}_{mn}F^{mn}$, typical for the classical axion electrodynamics. Since  the multiplier $\sin{\phi}$ in front of the pseudo-invariant $F^{*}_{mn}F^{mn}$ is the odd function, the second term remains a true scalar; the first term, which contains the even function $\cos{\phi}$ also is true scalar. As it was shown in \cite{B1,B2}, this idea of sin/cosine extension of the theory happened to be fruitful in application to cosmology, in particular, for description of axionically induced anomalous electric flares in the magnetized early Universe.

The next step in the extension of the axion theory was the study of  models nonlinear with respect to the Maxwell tensor. As an interesting case, we consider the nonlinear term ${\cal H}({\cal I})$ in the Lagrangian \cite{B2}, which tends to ${\cal I}$ for the small argument. In this work we follow this line, and formulate the relativistic axion magnetohydrodynamic nonlinear both in the axion field and in the Maxwell tensor. The work contains the general formalism of the relativistic nonlinear axion fluid dynamics, as well as, the truncated models of magnetohydrodynamics of two types. The model of the first type is based on the standard assumption that the electric conductivity $\sigma$ is very large, $\sigma \to \infty$; in this case one supposes that the electric field in the medium has to be vanishing providing the electric current to be finite. For the models of the second (principally new) type we assume that the axion field tends to the value $\phi \to \frac{\pi}{2}+ 2\pi n$ and the product $\sigma \cos{\phi} \to \infty \times 0$ remains finite, but the spatial part of the electric current now is vanishing. Such models are characterized by the anomalous growth of the axionically induced electric field at least in two cases: first, when there exists axionically induced magnetic conductivity; second, when the fluid flow possesses the rotation of the velocity four-vector. In this paper we formulate the general formalism and the perspective program of investigations, and in the nearest future we hope to apply the prepared formalism to the analysis of magnetohydrodynamic flows in cosmological and astrophysical systems.

\section{The formalism}

\subsection{The structure of the action functional}

The total action functional is considered to be presented by four elements
\begin{equation}
S_{(\rm tot)} = \int d^4x \sqrt{-g} \left\{ \frac{R+ 2\Lambda}{2\kappa} + {\cal L}_{(\rm EMA)} + L_{(\rm axion)} + L_{(\rm matter)}  \right\} \,.
\label{F0}
\end{equation}
Here $g$ is the determinant of the metric tensor; $R$ is the Ricci scalar; $\Lambda$ is the cosmological constant; $\kappa = 8 \pi G$ is the Einstein constant ($c=1$).
The Lagrangian of the electromagnetic field interacting nonlinearly with the pseudoscalar (axion) field $\phi$, indicated as ${\cal L}_{(\rm EMA)}$, is presented as an appropriate (linear or nonlinear) function
\begin{equation}
{\cal L}_{(\rm EMA)} = {\cal H}({\cal I})
\label{F1}
\end{equation}
of the unified invariant (\ref{000}). As usual,
$F_{mn}$ is the Maxwell tensor, and $F^{*mn} = \frac12 \epsilon^{mnpq}F_{pq}$ is its dual; the Levi-Civita tensor $\epsilon^{mnpq}= \frac{E^{mnpq}}{\sqrt{-g}}$ is defined with the equality $E^{0123}=1$. When the dimensionless pseudoscalar $\phi$ vanishes, we obtain from (\ref{000}) the standard invariant of the electromagnetic field $\frac14 F_{mn} F^{mn}$. When the pseudoscalar field is nonvanishing, but it tends to zero $\phi \to 0$, the unified invariant converts into  the term
\begin{equation}
{\cal I} \to  \frac14 \left[F_{mn} F^{mn} +  \phi \ F^{*}_{mn} F^{mn} \right] \,,
\label{F3}
\end{equation}
which is typical for the axion electrodynamics.
The Lagrangian of the pure pseudoscalar (axion) field
\begin{equation}
L_{(\rm axion)} = \frac12 \Psi^2_{0} \left[V(\phi) - \nabla_m \phi \nabla^m \phi \right]
\label{F4}
\end{equation}
contains the periodic axion potential
\begin{equation}
V(\phi) = 2m^2_{\rm A} \left(1- \cos{\phi}\right) \,,
\label{F5}
\end{equation}
which is invariant wit respect to the discrete symmetry transformation and converts into the potential $V=m^2_{\rm A} \phi^2 $, when $\phi$ is small. The parameter $m_{\rm A}$  describes the rest mass of the axion, and the parameter $\Psi_0$ is connected with the axion-photon coupling constant $g_{A \gamma \gamma}$ as follows $\frac{1}{\Psi_0}=g_{A \gamma \gamma}$.

The Lagrangian of the matter $L_{(\rm matter)}$ is not presented explicitly, and is the subject of phenomenological modeling.

\subsection{Master equations}

\subsubsection{Master equations for the electromagnetic field}

The Maxwell tensor $F_{ik}$ is connected with the potential of the electromagnetic field $A_k$ by the known relationship
\begin{equation}
F_{ik} = \nabla_{i}A_k - \nabla_k A_i \,.
\label{21}
\end{equation}
As the consequence of this definition one obtains the first series of the Maxwell equations
\begin{equation}
\nabla_k F^{*ik}=0 \,,
\label{20}
\end{equation}
which converts into identity, when (\ref{21}) holds. Variation of the total action functional (\ref{F0}) with respect to the electromagnetic potential $A_i$ gives the equations
\begin{equation}
\nabla_k H^{ik} = {\cal J}^i \,,
\label{F9}
\end{equation}
where
\begin{equation}
{\cal J}^i = - \frac{\delta L_{(\rm matter)}}{\delta A_i}
\label{F8}
\end{equation}
is the electric current, and
\begin{equation}
H^{ik} ={\cal H}^{\prime}({\cal I})\left[\cos{\phi} F^{ik} +
\sin{\phi} F^{*ik} \right]
\label{18}
\end{equation}
plays the role of the nonlinear tensor of the electromagnetic induction.
Since the following identity holds
\begin{equation}
\nabla_i \nabla_k H^{ik} = 0 \,,
\label{F10}
\end{equation}
one has to add the equation
\begin{equation}
\nabla_k {\cal J}^k = 0
\label{F11}
\end{equation}
into the total set of Master equations of the model. The term ${\cal J}^k$ is the subject of the phenomenological modeling.

\subsubsection{Master equation for the axion field}

Variation with respect to the pseudoscalar field $\phi$ yields
\begin{equation}
g^{mn} \nabla_m \nabla_n \phi + \frac12 \frac{dV}{d \phi} = - \frac{1}{ \Psi_0^2}\left\{ \frac14 {\cal H}^{\prime}({\cal I})\left[- \sin{\phi} \ F_{mn} F^{mn} +  \cos{\phi} \ F^{*}_{mn} F^{mn} \right] + {\cal G}  \right\} \,,
\label{F12}
\end{equation}
where the pseudoscalar source ${\cal G}$ appears formally as the variational derivative of the matter Lagrangian
\begin{equation}
{\cal G} = \frac{\delta L_{(\rm matter)}}{\delta \phi} \,.
\label{F13}
\end{equation}
The term ${\cal G}$ also is the subject of the phenomenological modeling.

\subsubsection{Master equations for the gravitational field}

Variation with respect to the metric gives the equations of the gravitational field
\begin{equation}
R_{pq} - \frac12 g_{pq} R - \Lambda g_{pq}= \kappa T^{(\rm tot)}_{pq} \,,
\label{F14}
\end{equation}
where $R_{pq}$ is the Ricci tensor. The total (effective) stress energy tensor $T^{(\rm tot)}_{pq}$ consists of three terms
\begin{equation}
T^{(\rm tot)}_{pq} = T^{(\rm EMA)}_{pq} + T^{(\rm axion)}_{pq} + T^{(\rm matter)}_{pq} \,.
\label{F15}
\end{equation}
The stress-energy tensor $T^{(\rm EMA)}_{pq}$, associated with the nonlinear electromagnetic field coupled  nonlinearly to the axion field, is of the form
\begin{equation}
T^{(\rm EMA)}_{ik} = {\cal H}^{\prime}({\cal I}) \cos{\phi} \left[\frac14 g_{ik} F_{mn}F^{mn} - F_{im}F_k^{\ m}\right] + g_{ik} \left[{\cal H}({\cal I}) - {\cal I} \cdot {\cal H}^{\prime}({\cal I}) \right] \,.
\label{17}
\end{equation}
It coincides with the standard stress-energy tensor of the electromagnetic field, when $\phi=0$ and ${\cal H}({\cal I})= {\cal I}$.
The trace of the tensor (\ref{17})
\begin{equation}
T^{(\rm EMA)}_{ik} g^{ik} = 4 \left[{\cal H}({\cal I}) - {\cal I} \cdot {\cal H}^{\prime}({\cal I}) \right]
\label{170}
\end{equation}
is not equal to zero, when we use the nonlinear version of the theory.

The stress energy tensor of the pure pseudoscalar field is presented as
\begin{equation}
T^{(\rm axion)}_{ik}= \Psi^2_0 \left\{\nabla_i \phi \nabla_k \phi  + \frac12 g_{ik}\left[V(\phi) - \nabla_p \phi \nabla^p \phi \right]\right\} \,.
\label{16}
\end{equation}
The stress-energy tensor of the matter is presented formally as
\begin{equation}
T^{(\rm matter)}_{pq} = \frac{(-2)}{\sqrt{-g}} \frac{\delta}{\delta g^{pq}} \left\{\sqrt{-g} L_{(\rm matter)}  \right\} \,.
\label{F151}
\end{equation}
It requires the algebraic decomposition and phenomenological decoding.

\subsubsection{Conservation law and balance equations}

The Bianchi identities require that
\begin{equation}
\nabla_{k} T^{(\rm tot)ik} =0 \,,
\label{F16}
\end{equation}
i.e., the total energy and momentum are conserved. In order to simplify the balance equations for the matter quantities, we present some auxiliary calculations. First, we consider the divergence of the axion stress-energy tensor (\ref{16}) on the solution to the equation (\ref{F12})
\begin{equation}
\nabla_k T^{ik (\rm axion)} = - \nabla^i \phi \left[\frac14 {\cal H}^{\prime}({\cal I})F^{mn} \left(-\sin{\phi}F_{mn} + \cos{\phi}F^*_{mn}  \right) + {\cal G}\right] \,,
\label{B1}
\end{equation}
second, we calculate the divergence of the electromagnetic stress-energy tensor (\ref{17}) on the solutions to the equations (\ref{F9}), (\ref{18})
\begin{equation}
\nabla_k T^{ik (\rm EMA)} = F^{ik}{\cal J}_k + \frac14 \nabla^i \phi  {\cal H}^{\prime}({\cal I})F^{mn} \left(-\sin{\phi}F_{mn} + \cos{\phi}F^*_{mn}  \right) \,,
\label{B2}
\end{equation}
and obtain finally
\begin{equation}
\nabla_k T^{ik (\rm matter)} = {\cal G} \nabla^i \phi - F^{ik} {\cal J}_k \,.
\label{B3}
\end{equation}

\subsection{Phenomenology}

\subsubsection{Macroscopic velocity four-vector and irreducible decomposition of its covariant derivative}

Phenomenological approach requires  the appropriate velocity four-vector $U^k$ to be defined as the starting point of the decomposition of the necessary quantities.
We follow the Eckart's approach \cite{Eckart} and consider the timelike unit velocity four-vector $U^k$ to be defined as follows:
\begin{equation}
N^k= n U^k  \,, \quad U^kU_k =1 \,, \quad n  = \sqrt{N_k N^k} = N^k U_k \,,
\label{F133}
\end{equation}
where  $N^k$ is the four-vector of particle number flux, and $n$ is the scalar of particle number. Generally, the plasma is the multi-component system, and thus $N^k = \sum\limits_{(a)}N^k_{(a)}$, where $(a)$ indicates the sort of particle.

With this four-vector we decompose all the tensor quantities using the so-called longitudinal and transversal components. In particular, the covariant derivative can be decomposed as follows
\begin{equation}
\nabla_k = U_k D + \nablab_k \,, \quad D = U^s \nabla_s \,, \quad \nablab_k = \Delta_k^j \nabla_j \,, \quad \Delta_k^j = \delta^j_k - U^j U_k \,.
\label{F134}
\end{equation}
$\Delta_k^j$ is the projector. The covariant derivative $\nabla_k U_j$ can be decomposed in the standard sum
\begin{equation}
\nabla_k U_j = U_k DU_j + \sigma_{kj} + \omega_{kj} + \frac13 \Delta_{kj} \Theta \,,
\label{F54}
\end{equation}
where the acceleration four-vector $DU_j$, the symmetric traceless shear tensor $\sigma_{kj}$, the skew - symmetric vorticity tensor $\omega_{kj}$ and the expansion scalar $\Theta$ are presented by the well-known formulas
\begin{equation}
DU_j = U^s \nabla_s U_j \,, \quad \sigma_{kj} = \frac12 \left(\nablab_k U_j + \nablab_j U_k \right) - \frac13 \Delta_{kj} \Theta \,, \quad \omega_{kj} = \frac12 \left(\nablab_k U_j - \nablab_j U_k \right) \,, \quad
\Theta = \nabla_kU^k \,.
\label{F541}
\end{equation}
The four-vector of the electric current also can be decomposed with respect to the $U^k$
\begin{equation}
{\cal J}^k = \rho U^k + \Jtr^k \,, \quad \rho = {\cal J}^m U_m
\,, \quad \Jtr^k = \Delta_k^j {\cal J}_j \,.
\label{F154}
\end{equation}

\subsubsection{Decomposition of the Maxwell tensor and its dual}

We use the standard definitions of the electric field four-vector $E^k$ and of the magnetic induction four-vector $B^k$ \cite{L1}
\begin{equation}
E^k = F^{km} U_m \,, \quad B^k = F^{*km} U_m \ \ \rightarrow E^k U_k = 0 \,, \quad B^k U_k = 0 \,,
\label{F542}
\end{equation}
which give the standard decompositions
\begin{equation}
F^{km} = E^m U^n - E^n U^m - \eta^{mns} B_s \,, \quad F^{*km} = B^m U^n - B^n U^m + \eta^{mns} E_s  \,.
\label{F542}
\end{equation}
Here the absolutely skew-symmetric symbol with three-indices is defined as $\eta^{mns}= \epsilon^{mnsl}U_l$; it is orthogonal to the velocity four-vector, $\eta^{mns}U_s=0$.

\subsubsection{Equation of the magnetic flux balance and the Faraday equation}

Now we apply the decompositions (\ref{F542}) to (\ref{20}) and consider the convolution $U_i \nabla_k F^{*ik}=0$. This procedure yields the scalar balance equation
\begin{equation}
\nablab_k B^k = \eta^{kmn}E_k \ \omega_{mn} \,.
\label{F551}
\end{equation}
Clearly, when the vorticity of the medium flow is absent, $\omega_{mn} = 0$, we deal with the standard conservation law of the magnetic flux.
Similarly, the convolution $\Delta_i^l \nabla_k F^{*ik}=0$ gives the Faraday law
\begin{equation}
\Delta^l_k D B^k + \eta^{lmn} \nablab_m E_n = - \frac23 \Theta B^l + B_k \left(\sigma^{kl} + \omega^{kl} \right) - \eta^{lmn} E_m D U_n + \Delta^l_s \epsilon^{smkn}  \omega_{kn} E_m \,,
\label{F552}
\end{equation}
the source terms in the right-hand side of this equation are produced by the non-uniformity and inhomogeneity of the medium flow.

\subsubsection{Axionic extension of the Gauss law for nonlinear electrodynamics}

The equations (\ref{F551}) and (\ref{F552}) do not contain information about the pseudoscalar (axion) field. The function $\phi$ and its gradient appears in the equations (\ref{F9}) due to the structure of (\ref{18}) and (\ref{000}). Convolution of (\ref{F9}) with the velocity four-vector gives the nonlinear axionic extension of the Gauss law
\begin{equation}
{\cal H}^{\prime}({\cal I}) \left\{ \cos{\phi} \left[\nablab_k E^k + \eta^{mpq} B_m \omega_{pq} + B^k \nablab_k \phi \right] - \sin{\phi}E^k \nablab_k \phi \right\} +
\label{G1}
\end{equation}
$$
+ {\cal H}^{\prime \prime}({\cal I})\left(E^k \cos{\phi} + B^k \sin{\phi} \right)\left\{\cos{\phi}\left[E_m B^m \nablab_k \phi + E^m \nablab_k E_m - B^m \nablab_k B_m \right] + \right.
$$
$$
\left. +\sin{\phi}\left[B^m \nablab_k E_m + E^m \nablab_k B_m + \frac12 \left(B^m B_m - E^m E_m \right) \nablab_k \phi \right]\right\} = -\rho \,,
$$
where $\rho = {\cal J}^m U_m$ is the charge density scalar. When $\phi=0$ and we deal with the linear electrodynamics, this equation reduces to the Gauss equation in the moving medium
\begin{equation}
\nablab_k E^k  = - \rho - \eta^{mpq} B_m \omega_{pq} \,.
\label{G11}
\end{equation}

\subsubsection{Axionic extension of the Ampere law for nonlinear electrodynamics}

Convolution of (\ref{F9}) with the projector $\Delta_i^l$ gives the equation, which can be indicated as the nonlinear axionic extension of the Ampere law
\begin{equation}
{\cal H}^{\prime}({\cal I}) \left\{ \cos{\phi} \left[\Delta^l_k DE^k {-} \eta^{lkm} \nablab_kB_m  {+} B^l D\phi {+} \eta^{klp} \nablab_k \phi E_p {+} \frac23 E^l \Theta {-} E_k (\sigma^{kl} {+} \omega^{kl}) {-} \eta^{lmn} B_m DU_n {-} \Delta^l_i \epsilon^{ikmn}B_m \omega_{kn} \right] {+} \right.
\label{G5}
\end{equation}
$$
\left. {+} \sin{\phi}\left({-} E^l D \phi {+} \eta^{lkp} B_p \nablab_k \phi \right)\right\} {+}
{\cal H}^{\prime \prime}({\cal I})\left[\left(E^l \cos{\phi} {+} B^l \sin{\phi} \right) D {\cal I} {+} \eta^{lkp}\left(\sin{\phi} E_p {-} \cos{\phi} B_p \right) \nablab_k {\cal I} \right] = \Jtr^l \,,
$$
where the following auxiliary notations are used:
\begin{equation}
D {\cal I} = \cos{\phi} \left[E^m DE_m - B^m D B_m + E_m B^m D\phi \right] + \sin{\phi} \left[\frac12 (B^m B_m - E^m E_m) D \phi + B^m DE_m + E^m DB_m \right]\,,
\label{G57}
\end{equation}
\begin{equation}
\nablab_k {\cal I} = \cos{\phi} \left[E^m \nablab_k E_m {-} B^m \nablab_k  B_m {+} E_m B^m \nablab_k  \phi \right] {+} \sin{\phi} \left[\frac12 (B^m B_m {-} E^m E_m) \nablab_k  \phi {+} B^m \nablab_k E_m {+} E^m \nablab_k B_m \right]\,.
\label{G58}
\end{equation}
When we deal with linear electrodynamics and $\phi=0$, we obtain from (\ref{G5}) the equation
\begin{equation}
     \Delta^l_k DE^k {-} \eta^{lkm} \nablab_kB_m  {+} \frac23 E^l \Theta {-} E_k (\sigma^{kl} {+} \omega^{kl}) {-} \eta^{lmn} B_m DU_n {-} \Delta^l_i \epsilon^{ikmn}B_m \omega_{kn} = \Jtr^l \,,
\label{G599}
\end{equation}
which can be indicated as the Ampere equation in the moving medium.

\subsubsection{Decomposition of the electric current}

The equation $\nabla_k {\cal J}^k = 0$ can be now rewritten as
\begin{equation}
D \rho + \rho \Theta  = \Jtr^k DU_k  - \nablab_k \Jtr^k \,,
\label{F154}
\end{equation}
and can be considered as the evolutionary equation for the charge density $\rho$.
In the standard relativistic magnetohydrodynamics the transversal component of the current four-vector is of the form
\begin{equation}
\Jtr^k_{(\rm standard)} = \sigma E^k \,,
\label{F138}
\end{equation}
where $\sigma$ is the conductivity scalar.
In general case we present the  transversal component of the current four-vector as the series
\begin{equation}
\Jtr^k =  {\cal J}^k_{(1)} + {\cal J}^k_{(2)}  + ...
\label{F23}
\end{equation}
with respect to the number of derivatives, as the Effective Field Theory advises \cite{EFT}. In this sense the term ${\cal J}^k_{(1)}$ contains only one derivative of the first order; the term ${\cal J}^k_{(2)}$ contains the composition of two derivatives of the first order; second derivatives are omitted. We restrict our-selves by the first order terms, and obtain only five appropriate ones
\begin{equation}
\Jtr^k = \sigma E^k \cos{\phi}  + \tilde{\sigma} B^k \sin{\phi}   + \nu_1 \sin{\phi} \nablab^k \phi + \nu_2 \cos{\phi} D U^k + \nu_3 \eta^{kpq} \sin{\phi}\ \omega_{pq}    \,.
\label{F24}
\end{equation}
The construction $\tilde{\sigma} \sin{\phi}$ plays the role of a magnetic conductivity associated with the chirality introduced by the axion field into the electrodynamic system; this term describes the current directed along the magnetic induction four-vector. The term  $\nu_2 \cos{\phi} D U^k$ has the classical analog; it describes the electric current caused by the acceleration of the conductor.  The term $\nu_1 \sin{\phi} \nablab^k \phi$ relates to current along the gradient of the axion field. The last term in (\ref{F24}) rewritten as $2\nu_3   \omega^k \sin{\phi}$ with the help of the  angular velocity four-vector $\omega^k = \frac12 \eta^{kpq}  \omega_{pq}$ can be attributed to the current provoked by the rotation of the chiral medium.

\subsubsection{Modification of the equation of the axion field}

Taking into account the representation of the electromagnetic equations we can now rewrite the equation for the axion field as follows:
\begin{equation}
g^{mn} \nabla_m \nabla_n \phi + m^2_{A} \sin{\phi} = - \frac{1}{ \Psi_0^2}\left\{  {\cal H}^{\prime}({\cal I})\left[ \frac12 \sin{\phi}  (B^m B_m -E_m E^m ) +  \cos{\phi} E^m B_m \right] + {\cal G}  \right\} \,.
\label{0F12}
\end{equation}
The pseudoscalar source ${\cal G}$ can be phenomenologically decomposed similarly to the electric current four-vector; recollecting the terms of the zero, first and second order in derivatives we  obtain
\begin{equation}
{\cal G} =\omega_1 \Theta \sin{\phi}
+\omega_2 \cos{\phi} D \phi
+\omega_3 \Theta \cos{\phi} D \phi
+\omega_4 DU^k \cos{\phi}\nablab_k \phi +
\label{F149}
\end{equation}
$$
+\omega_5 E^k \cos{\phi} \nablab_k \phi
+\omega_6 \sin{\phi} E^k DU_k
+\omega_7 \cos{\phi} \eta^{kpq} E_k \omega_{pq}
+
$$
$$
+\omega_8 \cos{\phi }B^k DU_k
+\omega_9 \sin{\phi} B^k \nablab_k \phi
+\omega_{10}\cos{\phi} \eta^{kpq} B_k  \omega_{pq} \,.
$$

\subsubsection{Decomposition of the stress - energy tensor of the matter and the evolutionary equation for the velocity four-vector}

The algebraic decomposition of the stress-energy tensor of the matter is well known:
\begin{equation}
T^{ik(\rm matter)}
= W U^i U^k + U^i q^k + U^k q^i - \Delta^{ik} P + \Pi^{ik} \,,
\label{03}
\end{equation}
where $W$ is the scalar of the matter energy density, $q^i$ is the heat-flux four-vector, $P$ is the Pascal equilibrium pressure, and $\Pi^{ik}$ is the tensor of non-equilibrium pressure.
The rate of evolution of the scalar $W$, i.e., the quantity $DW$, can be found from the equation $U_i \nabla_k T^{ik(\rm total)}=0$ accounting for (\ref{B3}):
\begin{equation}
DW + (W+P) \Theta = q^k DU_k - \nabla_k q^k + \Pi^{ik}\left(\sigma_{ik}+ \frac13 \Delta_{ik} \Theta \right) + {\cal G} D \phi + E_k \Jtr^k \,.
\label{B12}
\end{equation}
Convolution of (\ref{B3}) with the projector $\Delta^l_i$ gives the equation for the macroscopic velocity of the medium dynamics
\begin{equation}
(W+P) DU^l  = \nablab^l P - \Delta^l_i Dq^i - q^l \Theta - q^k \nablab_k U^l - \Delta^l_i \nabla_k \Pi^{ik}  - \rho E^l + {\cal G} \nablab^l \phi + \eta^{lks} B_s \Jtr_k \,.
\label{B121}
\end{equation}
Let us add that if we follow the Eckart's approach we have to keep in mind that the heat-flux four-vector
\begin{equation}
q^i = \lambda \left(\nablab^i T {-} T DU^i \right)
\label{Eck1}
\end{equation}
contains the phenomenological constant $\lambda$, describing the heat conductivity, the spatial gradient of the temperature $T$ and the acceleration four-vector $DU^k$. As for the anisotropic pressure tensor, which satisfies the relationships
\begin{equation}
\Pi_{ik} = \Pi_{(0)ik} + \Pi \Delta_{ik} \,, \quad \Pi_{(0)ik} g^{ik} = 0 \,, \quad \Pi = \frac13 \Pi_{ik} g^{ik} \,,
\label{Eck2}
\end{equation}
it can be presented using two phenomenological constants: the shear viscosity $\eta$ and bulk viscosity $\zeta$, as follows:
\begin{equation}
\Pi_{ik(0)} = \eta \sigma_{ik} \,, \quad \Pi = 3 \zeta \Theta  \,.
\label{Eck3}
\end{equation}
Also, we assume that $W$ and $P$ are connected by the two-parameter equation of state
\begin{equation}
W = W(n,T) \,, \quad P = P(n,T) \,,
\label{Eck4}
\end{equation}
and by the compatibility condition
\begin{equation}
n \frac{\partial W}{\partial n} + T \frac{\partial P}{\partial T} = W+P \,.
\label{Eck5}
\end{equation}
In particular case, when $W$ is linear in $n$, i.e., $W=n e(T)$, we obtain immediately from (\ref{Eck5}) that $P=f(n) T$ with arbitrary function $f(n)$, and we can extract the known equation for the pressure of the relativistic perfect gas $P= nk_B T$ ($k_B$ is the Boltzmann constant).

At the end of this Section we would like to emphasize that till now we considered the general formalism, which is appropriate for moving electromagnetically active fluid. In the next Section we start to discuss  magnetohydrodynamic models with specific ansarz concerning the electric conductivity of the system.

\section{Two examples of truncated sets of equations of the axion magnetohydrodynamics}

\subsection{Classical approach: Approximation of infinite electric conductivity}

\subsubsection{Auxiliary equations}

Zero order approximation of classical magnetohydrodynamics is based on the assumption that $\sigma \to \infty$. In this situation one assumes that the electric field four-vector has to tend to zero, $E^k \to 0$ providing the product $\sigma E^k$ remains finite.
In fact, we have to decompose the four-vector of the electric field in the power series with respect to small parameter $\frac{1}{\sigma}$, and in the zero order approximation to put $E^i =0$  in all the Master equations. As for the Ampere equation  (\ref{G5}), it converts now into the equation for the electric field
$$
E^k  = \frac{1}{\sigma  \cos{\phi}}\left\{ - \tilde{\sigma} B^k \sin{\phi}   - \nu_1 \sin{\phi} \nablab^k \phi - \nu_2 \cos{\phi} D U^k - \nu_3 \eta^{kpq} \sin{\phi}\ \omega_{pq}  + \right.
$$
\begin{equation}
\left. + {\cal H}^{\prime}({\cal I}) \left[ \cos{\phi} \left( {-} \eta^{lkm} \nablab_kB_m  {+} B^l D\phi  {-} \eta^{lmn} B_m DU_n {-} \Delta^l_i \epsilon^{ikmn}B_m \omega_{kn} \right) {+}  \sin{\phi} \ \eta^{lkp} B_p \nablab_k \phi \right]+ \right.
\label{electric}
\end{equation}
$$
\left. {+} \frac12 {\cal H}^{\prime \prime}({\cal I})\left\{ B^l \sin{\phi} \left[\sin{\phi}  B^m B_m D \phi {-} \cos{\phi}  D (B^m B_m)\right] {-} \eta^{lkp}B_p \cos{\phi}  [\sin{\phi} B^m B_m \nablab_k \phi\ {-} \cos{\phi}  \nablab_k (B^m B_m )] \right\} \right\}\,.
$$
The Gauss  equation (\ref{G1}) can be now considered as the definition of the charge density scalar \begin{equation}
- \rho = {\cal H}^{\prime}({\cal I}) \cos{\phi} \left(\eta^{mpq} B_m \omega_{pq} + B^k \nablab_k \phi \right)  +
\frac12 {\cal H}^{\prime \prime}({\cal I})B^k \sin{\phi} \left[\sin{\phi} B^m B_m  \nablab_k \phi -\cos{\phi} \nablab_k (B^m B_m) \right] \,.
\label{0G177}
\end{equation}

\subsubsection{Equations for the magnetic field}

The equations (\ref{F551}) and (\ref{F552}) can be now written in the form
\begin{equation}
\nablab_k B^k = 0\,,
\label{01}
\end{equation}
\begin{equation}
\Delta_{kl} D B^k  =  B^k \nablab_k U_l - B_l \Theta \,.
\label{02}
\end{equation}
If we consider the case, when the particle number $n$ is conserved and thus
\begin{equation}
\nabla_k (n U^k) = 0  \ \  \rightarrow Dn + n \Theta = 0 \,,
\label{027}
\end{equation}
we can rewrite the equation (\ref{02}) in the form
\begin{equation}
\Delta^l_{k} {\pounds}_U \left(\frac{B^k}{n}\right) = 0 \,,
\label{029}
\end{equation}
where ${\pounds}_U$ is the Lie derivative calculated along the four-vector $U^j$. The equation (\ref{029}) is known as the condition of frostbite of magnetic field lines. Formally speaking, this condition does not depend on the axion field and does not include information about the nonlinearity of the axion electrodynamics.

\subsubsection{Equation for the axion field}

The Master equation for the axion field is
\begin{equation}
g^{mn} \nabla_m \nabla_n \phi +  \sin{\phi} \left[m^2_{A} + \frac{1}{ 2\Psi_0^2}{\cal H}^{\prime}({\cal I}) B^m B_m  \right]=
\label{1axion00}
\end{equation}
$$
- \frac{1}{ \Psi_0^2}\left[\cos{\phi}\left(\omega_2 D \phi +\omega_3 \Theta D \phi +\omega_4 DU^k \nablab_k \phi +\omega_8 B^k DU_k +\omega_{10} \eta^{kpq} B_k  \omega_{pq} \right) +
\sin{\phi}\left(\omega_1 \Theta +\omega_9 B^k \nablab_k \phi  \right) \right]\,.
$$

\subsubsection{Equations for the velocity four-vector}

The equations for the velocity four-vector (\ref{B121}) can be reconstructed as follows: we have to work with the equation
\begin{equation}
(W+P) DU^l  = \nablab^l P {-} \lambda \Delta^l_i D(\nablab^i T {-}T DU^i) {-} \lambda \Theta (\nablab^lT {-} T DU^l) {-}  \lambda (\nablab^kT {-} T DU^k) \nablab_k U^l {-} \Delta^l_i \nabla_k \Pi^{ik}  {+} {\cal G} \nablab^l \phi {+} \eta^{l}_{\cdot js} B^s \Jtr^j  \,,
\label{B121j}
\end{equation}
where $\Jtr^j$ should be replaced by the term
\begin{equation}
\Jtr^j = {\cal H}^{\prime}({\cal I}) \left[ \cos{\phi} \left({-} \eta^{jkm} \nablab_kB_m  {+} B^j D\phi    {-} \eta^{jmn} B_m DU_n {-} \Delta^j_i \epsilon^{ikmn}B_m \omega_{kn} \right) {+}  \sin{\phi} \eta^{jkp} B_p \nablab_k \phi \right] +
\label{G5j}
\end{equation}
$$
{+}
\frac12 {\cal H}^{\prime \prime}({\cal I})\left\{B^j \sin{\phi} \left[\sin{\phi}  B^m B_m  D \phi - \cos{\phi} D (B^m B_m) \right] {-} \eta^{jkp} \cos{\phi} B_p  \left[\sin{\phi}  B^m B_m  \nablab_k  \phi - \cos{\phi} \nablab_k (B^m  B_m ) \right] \right\}  \,,
$$

\subsubsection{The case of linear electrodynamics}

When ${\cal H}({\cal I}) = {\cal I}$, we can simplify the  equations (\ref{0G177}) and (\ref{electric}) as follows:
\begin{equation}
\rho = - \cos{\phi} \left(\eta^{mpq} B_m \omega_{pq} + B^k \nablab_k \phi \right) \,,
\label{0G1}
\end{equation}
$$
E^k  = \frac{1}{\sigma }\left[\tan{\phi} \left(  \eta^{lkp} B_p \nablab_k \phi - \tilde{\sigma} B^k    - \nu_1  \nablab^k \phi  - \nu_3 \eta^{kpq}  \omega_{pq} \right) + \right.
$$
\begin{equation}
\left. +   \left( B^l D\phi - \nu_2 D U^k  {-} \eta^{lkm} \nablab_kB_m    {-} \eta^{lmn} B_m DU_n {-} \Delta^l_i \epsilon^{ikmn}B_m \omega_{kn} \right)   \right] \,.
\label{electric2}
\end{equation}

\subsection{New versions of axion magnetohydrodynamics}

\subsubsection{Anomalous regime in the presence of axionically induced magnetic conductivity}

As was shown in \cite{B1,B2} the nonlinear axion electrodynamics admits the existence of a anomalous regime, which is characterized by the electric field $E^k \propto \tan{\phi} B^k$. When $\phi \to \frac{\pi}{2} + 2\pi n$, the electric field grows infinitely. In the model under consideration we see that the term $\sigma \cos{\phi} E^k$ in (\ref{F24})
remains finite at $\sigma \to \infty$, if $\cos{\phi} \to 0$, i.e., $\phi \to \frac{\pi}{2} + 2\pi n$. In other words, we can remove the requirement that $E^k \to 0$, and the electric field remains finite. There are two interesting phenomenological situations based on this idea. First, when $\tilde{\sigma} \neq 0$, but $\nu_3=0$, we can connect the electric and magnetic field by the relationship
\begin{equation}
E^k = -\frac{\tilde{\sigma}}{\sigma}  B^k \tan{\phi} \,,
\label{electric5}
\end{equation}
keeping in mind that the big value of conductivity parameter in the denominator  is compensated by the large value of the function $\tan{\phi}$.
In order to obtain the corresponding truncated set of equations we can put
\begin{equation}
\phi \to \frac{\pi}{2} - \psi \,, \quad |\psi|<<1 \,, \quad  \sin{\phi} \to \cos{\psi}  \,, \quad \cos{\phi} \to \sin{\psi}\,, \quad \tan{\phi} \to \cot{\psi} \,, \quad
E^k \to  -\frac{\tilde{\sigma}}{\sigma}  B^k \cot{\psi} \to -\frac{\tilde{\sigma}}{\sigma \psi}  B^k  \,,
\label{psi1}
\end{equation}
to all the equations obtained in Section II. One can present the result of this procedure as follows.
The equation (\ref{F551}) converts into
\begin{equation}
\nablab_k B^k = -\frac{\tilde{\sigma}}{\sigma \psi} \eta^{kmn}B_k \ \omega_{mn} \,,
\label{psi2}
\end{equation}
the Faraday law (\ref{F552}) takes the form
\begin{equation}
\Delta^l_k D B^k + \Theta B^l - B^k \nablab_k U^l  = \frac{\tilde{\sigma}}{\sigma \psi} \left[ \eta^{lmn} \nablab_m B_n +  \eta^{lmn} B_m  D U_n  - \Delta^l_s \epsilon^{smkn}  \omega_{kn} B_m  + \eta^{lmn} B_n \frac{ \nablab_m \psi}{\psi} \right]\,.
\label{psi3}
\end{equation}
The leading order version of the axionically modified Gauss law (\ref{G1}) can be now written as
\begin{equation}
B^k \nablab_k \left[{\cal H}^{\prime}({\cal I})\right] = - \rho \,,  \quad  {\cal I} = - \frac{\tilde{\sigma}B^m B_m}{\sigma \psi}  \,.
\label{psi6}
\end{equation}
The leading order version of the axionically modified Ampere law (\ref{G5}) can be presented in the form
\begin{equation}
B_p \left(g^{lp} D + \frac{\tilde{\sigma}}{\sigma \psi}\eta^{lpk} \nablab_k \right) {\cal H}^{\prime}({\cal I}) = 0 \,.
\label{psi9}
\end{equation}
In the leading order approximation with respect to $\psi$ the equation (\ref{0F12}) for the axion field requires the function $\psi$ to be found from the equation
\begin{equation}
  \Psi_0^2 m^2_{A} + \omega_1 \Theta  = {\cal I} {\cal H}^{\prime}({\cal I}) \frac{\sigma \psi}{2\tilde{\sigma}}  \left[1 - \left(\frac{\tilde{\sigma}}{\sigma \psi} \right)^2 \right]  - \frac{\omega_6 \tilde{\sigma}}{\sigma \psi} B^k DU_k  \,.
\label{psi11}
\end{equation}
In particular, when $\omega_1=\omega_6 =0$, and the function ${\cal H}({\cal I})$ is logarithmic, i.e., ${\cal H}({\cal I}) = \frac{1}{\nu} \log{\frac{{\cal I}}{{\cal I}_*}}$ ($\nu$ and ${\cal I}_*$ are some constants) we see that $\psi$ takes constant value
\begin{equation}
 \psi = \frac{\tilde{\sigma}}{\sigma} \left[\nu \Psi^2_0 m^2_{A} \pm \sqrt{\nu^2 \Psi^4_0 m^4_{A}+1} \right] \,.
\label{psi12}
\end{equation}
The final remark concerns the equation for the velocity four-vector (\ref{B121}). Since now $\Jtr^k=0$, this equation transforms into
\begin{equation}
(W+P) DU^l  - \nablab^l P + \Delta^l_i Dq^i + q^l \Theta + q^k \nablab_k U^l + \Delta^l_i \nabla_k \Pi^{ik} =
\label{B121}
\end{equation}
$$
=  - \frac{\tilde{\sigma}}{\sigma \psi} B^l B^k \nablab_k \left[{\cal H}^{\prime}({\cal I})\right]  + \left(\omega_1 \Theta - \frac{\omega_6 \tilde{\sigma}}{\sigma \psi} B^k DU_k \right)\nablab^l \phi  \,.
$$

\subsubsection{Anomalous regime in the presence of rotation in the magnetohydrodynamic flow}

The second interesting case appears, if $\tilde{\sigma}=0$, $\nu_3 \neq 0$ and there exists the rotation of the medium, i.e.,
the angular velocity of the medium flow rotation is nonvanishing, $\omega^k = \frac12 \eta^{kpq} \omega_{pq} \neq 0$. The most interesting application of this model would be the model of plasma dynamo. Based on the arguments presented above, in this second case we can put (instead of (\ref{electric5}))
\begin{equation}
E^k = -\frac{2\nu_3}{\sigma} \omega^{k}  \tan{\phi} \to -\frac{2\nu_3}{\sigma \psi} \omega^{k}
\label{electric6}
\end{equation}
into the axionically extended Faraday, Gauss and Ampere equations, as well as, into the equations for the pseudoscalar field and for the velocity four vector.
This procedure is analogous to the one presented in the previous subsection.

\section{Outlook}

In this work we presented the mathematical formalism of the relativistic nonlinear axion magnetohydrodynamics. The term nonlinear means that both fields: the pseudoscalar (axion) and the electromagnetic one are described in the nonlinear version. In fact, we entered the threshold of the new program execution, which can be formally divided into two sectors.

\noindent
1. In the first sector of the future work we plan to extend the general formalism; to be more precise we plan to do the following.

\noindent
1.1.  We plan to consider the  axionically extended  Born-Infeld, Euler-Heisenberg, etc. models, which are nonlinear in the first and second invariants of the electromagnetic field.

\noindent
1.2. We plan to supplement the hydrodynamic part of the established theory, which is now based on the Eckart approach (models with bulk and shear viscosity and heat conductivity), by the elements of the Israel-Stewart theory \cite{IS}, which deals with transient irreversible (second order) relativistic thermodynamics.

\noindent
1.3. We plan to prepare the justification of the nonlinear axion magnetohydrodynamics on the base of extension of the relativistic kinetic theory of the axionically active plasma, using the axionic modification of the Lorentz force.

\noindent
2. In the second sector of the future work we plan to study the anomalous regimes of the magnetohydrodynamic flows; for this purpose we hope to do the following.

\noindent
2.1. We hope to consider new models of anomalous accretion of the magnetized matter on the rotating neutron stars, black holes and dyons, as well as instabilities of a new type caused by the interactions with the axion field.

\noindent
2.2. We hope to study new anomalous models of dynamo, jets production and turbulent flows.

\noindent
2.3. We hope to analyze the solutions describing magnetohydrodynamic waves, shock waves based on the introduction of the correspondingly generalized Reynolds numbers.

Wish success to our group.

\acknowledgments{The work was supported by Russian Foundation for Basic Research (Grants No 20-02-00280 and 20-52-05009).}

\end{document}